\documentclass[aps,prl,preprint,groupedaddress,nofootinbib,longbibliography]{revtex4}
\usepackage{amsmath, latexsym, amsfonts, amssymb, amsthm, amscd}
\usepackage{graphicx, color}

\newcommand{\be}{\begin{equation}} 
\newcommand{\ee}{\end{equation}}

\begin{document}

\title{We've walked a million miles for one of these smiles}

\author{L. De Leo, V. Vargas, S. Ciliberti, J.-P. Bouchaud}
\affiliation{Capital Fund Management, 6 Bd Haussmann, 75009 Paris}

\begin{abstract}
We derive a new, exact and transparent expansion for option smiles,
which lends itself both to analytical approximation and to congenial numerical treatments. 
We show that the skew and the
curvature of the smile can be computed as exotic options, for which the Hedged
Monte Carlo method is particularly well suited. When applied to options on the
S\&P index, we find that the skew and the curvature of the smile are very
poorly reproduced by the standard Edgeworth (cumulant) expansion. Most notably, 
the relation between the skew
and the skewness is inverted at small and large vols, a feature that none of
the models studied so far is able to reproduce. Furthermore, the
around-the-money curvature of the smile is found to be very small, in stark
contrast with the highly kurtic nature of the returns.
\end{abstract}


\maketitle

Understanding the shape of volatility smiles in option markets is arguably one
of the most active field of research in quantitative finance \cite{Gatheral,
Sircar}. By definition, the existence of an option smile is the sign that the
standard Black-Scholes model is not an adequate representation of the
stochastic dynamics of financial assets. A huge variety of models have been
proposed over the years in order to account for the non-Gaussian nature of
price changes and the corresponding option smiles: jumps and L\'evy processes
\cite{Tankov}, GARCH and stochastic volatility models \cite{PHL} (including the
popular Heston and SABR models \cite{SABR}), multiscale (multifractal) models
\cite{BMD,BMD2}, mixed jumps/stochastic vol. models, etc. Another model-free strand
of research that sheds light on the origin and the general structure of option
smiles, is to assume that the corrections to the Black-Scholes model can
somehow be considered as ``small''. A natural idea is to use a Edgeworth
cumulant expansion of the distribution of the price change
$r_T=\frac{S_T-S_0}{S_0}$ (see Appendix). Working with the Bachelier model as
benchmark (i.e. $(S_t)_{t \geq 0}$ follows a Brownian motion instead of a
Geometric Brownian motion), the authors of \cite{cf:BoCoPo,cf:BoPo} have
obtained the following cumulant expansion for the smile:
\begin{equation}\label{eq:smile1}
    \sigma_{BS}=\sigma (1+\frac{\mathcal{S}_T}{6} \mathcal{M} +
    \frac{\kappa_T}{24}(\mathcal{M}^2-1)+ \dots),
    \qquad \mathcal{M}:=\frac{K-S_0}{S_0\sigma \sqrt{T}}
\end{equation}
where $\sigma\sqrt{T}$ denotes the standard deviation of $r_T$, $\mathcal{M}$
is the rescaled moneyness of the option, and $\mathcal{S}_T\ll 1$ and
$\kappa_T\ll 1$ are respectively the skewness and kurtosis of $r_T$, which are
assumed to be small for the expansion to make sense.\footnote{Eq.
(\ref{eq:smile1}) in fact assumes that $\mathcal{S}_T^2\ll\kappa_T$, which is
often the case in practice.} Note that the authors of \cite{cf:BaFo}
independently obtained a similar formula within a Black Scholes context, around
the same time. The two formulas coincide for $\sigma \sqrt{T}\ll1$, which is
expected because in that limit the Black-Scholes model becomes identical to the
Bachelier model. 

The interest of the above formula is that it allows one to understand why
smiles have generically an asymmetric parabolic shape, with an asymmetry
related to the skewness of the distribution of the underlying, while the
curvature of the smile is proportional to its kurtosis. Since from general
arguments the distribution of $r_T$ becomes Gaussian at large times (albeit
perhaps very slowly due to long-memory effects \cite{cf:BoPo}), the asymmetry
and curvature of the smile are expected to go down with maturity $T$, as indeed
observed in option markets.

However, the expansion (\ref{eq:smile1}) involves moments up to order $4$ and
as such is troublesome both theoretically and practically. Indeed, the
cumulative distribution of returns $P(|r_T|> x)$ is found empirically to
decrease as $x^{-\mu}$ with $\mu \approx 3$ for many assets (see e.g.
\cite{cf:GaSt,cf:GaSt2}) and thus the moment of order $3$ is formally divergent. At any
rate, it is in practice very difficult to estimate moments of order $3$ and $4$
because of the noise induced by large events. One possibility is to replace the
skewness and kurtosis of $r_T$ by some lower moment approximations (cf.
\cite{cf:BoPo}). However, this procedure is ambiguous as there is freedom in
the choice of lower moments to be used.

The purpose of this note is to establish a different smile expansion formula,
which is general, rigorous and involves no moments of order greater than $2$.
This new smile formula lends itself to analytical treatments, which allow one
to recover, for example, the recent results of Bergomi \& Guyon \cite{cf:BeGu}.
But perhaps more importantly, our formula can be coupled to the ``Hedged
Monte-Carlo Method'' of \cite{cf:HMC,cf:HMC2} to yield a powerful numerical method,
which provides accurate estimates of the smile parameter for arbitrarily complex
models of the underlying. One may even use historical data directly,
short-circuiting any modelling assumption. Our formula, derived in the
Appendix, reads: 
\begin{equation}\label{eq:smile2} 
\sigma_{BS}=\sigma (\alpha_T+\beta_T \mathcal{M}+\gamma_T \mathcal{M}^2+ O(\mathcal{M}^3)),
\end{equation}
with the coefficients $\alpha_T, \beta_T, \gamma_T$ given by:
\begin{equation}
\alpha_T=\sqrt{\frac{\pi}{2}}E[|u_T|],\quad
\beta_T= \sqrt{\frac{\pi}{2}}\left[1-2P(u_T>0)\right], \quad
\gamma_T= \sqrt{\frac{\pi}{2}}\, p_T(0)-\frac{1}{2\alpha_T},
\end{equation}
where $u_T=r_T/\sigma\sqrt{T}$ and $p_T(.)$ is the density of $u_T$. Using a
Edgeworth expansion, one can show that both smile formulae (\ref{eq:smile1})
and (\ref{eq:smile2}) coincide in the limit where cumulants are small and
cumulants higher than four can be neglected (see Appendix). One finds in
particular $\beta_T \approx \mathcal{S}_T/6$. But what is remarkable is that
while the ``old'' smile formula (\ref{eq:smile1}) is highly sensitive to
extreme events, the new formula (\ref{eq:smile2}) only involves low moments of
the distribution of $u_T$. In particular, the skew of the smile, as measured by
the coefficient $\beta_T$, is technically a moment of order $0$, i.e. the large
events do not play any role at all.

This new smile formula can be used in different ways. One can for example
obtain directly the coefficients $\alpha_T, \beta_T, \gamma_T$ in a vol-of-vol
expansion, recovering the Bergomi-Guyon results to lowest order \cite{Vargas}. 
One of the salient results of Bergomi \& Guyon \cite{cf:BeGu} is
that to leading order in vol-of-vol and for a broad family of linear models,
the Edgeworth expansion result $\beta_T = \mathcal{S}_T/6$ holds, i.e. the {\it
skew} of the smile $\beta$, and the {\it skewness} $\mathcal{S}_T$ of the
distribution of $u_T$, are very simply related. A first exercise is to use our
formula to investigate the case where the correlation between stock returns and
volatility is non-linear. Assume for example a stochastic volatility model
where $r_T=\int_0^T \sigma_t dW_t$, with $dW_t$ the standard Wiener process,
and where the instantaneous volatility $\sigma_t^2$ is given by:
\begin{equation}\label{eq:non-linear}
    \sigma_t^2= \overline{\sigma}^2 \left[1 + 2\epsilon(|\xi_t+\theta|\, 
    1_{\xi_t+\theta<0})\right],
    \qquad \xi_t=\omega \int_{0}^t e^{-\omega (t-u)}dW_u, \quad \epsilon > 0.
\end{equation}
Here $\overline{\sigma}^2$ is a ``baseline'' volatility (which might itself 
evolve over long times scales, see below).

When the threshold $\theta$ is zero, the above equation means that when recent
returns are negative, the future volatility will be higher than usual (this is
the standard leverage effect \cite{cf:BoMaPo,cf:BoPo}), while when the recent
returns are positive, there is no impact on the future vol. When $\theta > 0$,
only relatively large negative returns will increase the volatility, while even 
small positive returns increase the volatility when $\theta < 0$.
The coefficients $\beta_T$ and $\mathcal{S}_T$ can be
computed to first order in $\epsilon$ in this case \cite{Vargas}. One finds
that the equality $\beta_T = \mathcal{S}_T/6$ only holds when $\theta=0$. When
$\theta > 0$, one the other hand, the skew amplitude $|\beta_T|$ is {\it
smaller} than $|\mathcal{S}_T|/6$, and vice-versa when $\theta < 0$. This shows
that even for small vol-of-vols, non linear effects can significantly affect
the relation between skew and skewness, and that there can be no general link
between the two.

Although analytical results are interesting, we believe that the true
added-value of our new smile expansion comes from the following remark: the
three coefficents $\alpha_T, \beta_T$ and $\gamma_T$ can be interpreted as the
{\it average payoff of some ``exotic'' options}. Indeed, $\alpha_T$ is simply
an at-the-money straddle, $\beta_T$ is related to an at-the-money binary
option, and the first term of $\gamma_T$ is a ``no move'' option which pays if
the underlying ends very close to its initial price. \footnote{It is in fact common folklore,
inherited from Eq. (\ref{eq:smile1}), that buying slightly out of the money and selling slightly in 
the money allows one to bet on the ``skewness'', while a butterfly trade allows one 
to bet on the ``kurtosis''.}  Having interpreted these
coefficients as option prices, one can use the ``Hedged Monte-Carlo'' (HMC)
method proposed in \cite{cf:HMC,cf:HMC2} to price these options numerically. This is
interesting for at least two reasons (see \cite{cf:HMC,cf:HMC2,cf:BoPo,cf:Kapoor}
for a more thorough discussion): first, the HMC has by construction a low
variance, that allows one to price options accurately with a relatively small
number of paths; second, the P\&L of the hedge automatically transforms
historical or empirical probabilities into risk-neutral ones.  Technically, the
pricing of the ``exotic'' options is done in a way very similar to what is
described in \cite{cf:HMC,cf:HMC2}, except that we do not try here to determine the
optimal hedge for these options, but rather use a (sub-optimal) Black-Scholes
hedge. This reduces HMC to a standard reduced-variance Monte-Carlo \cite{cf:Clewlow} 
which is conceptually 
much simpler and faster numerically. Although only approximate, this pruned down version of HMC 
is accurate enough for the present purpose. The other subtle point concerns the
pricing of the ``no move'' option, since one has to introduce a window with a
small but finite width $\delta$, and carefully extrapolate the results to
$\delta = 0$. We in fact use a Gaussian payoff function,
$\exp(-u_T^2/2\delta^2)/\sqrt{2 \pi \delta^2}$ for that purpose.  We have
calibrated the method on known cases -- for example on the above non-linear
model, with good agreement between the analytical calculations and the HMC
results. 

We now apply our method to real time series, using historical trajectories to
price our exotic options and determine directly the skew $\beta_T$ and
curvature $\gamma_T$ of the smile, without any a priori model. The data we use
is the time series of daily returns (open-high-low-close) of the S\&P500 index in the period
1/1/1970 -- 31/12/2011. We divide the sample into two bins: one corresponding to
high volatilities (larger than the median), the other to low volatilities,
where the volatility is a 20 day exponential moving average of a Rogers-Satchell estimator
of the squared daily volatility {\it before} the day the smile coefficients are determined. We
show in Fig. 1 the ``fair'' skew $\beta_T$ as a function of $T$ for options
between 1 and 20 days, that we compare with the prediction $\mathcal{S}_T/6$
based on a direct measure of the third moment skewness of the distribution of $u_T$. 
We find,
perhaps surprisingly, that while $|\mathcal{S}_T|/6$ is systematically larger than
the skew $|\beta_T|$ for large volatilities, the opposite is true for low
volatilities. Let us emphasize here that we are not speaking of implied 
volatility smiles from option markets, but rather of theoretical predictions of 
what the {\it fair} parameters of the smile should be. Including some information 
from option markets could be done along the lines of \cite{Avellaneda}.

In Fig. 2, we show the smile curvature $\gamma_T$ as a function of $T$ for the
same two volatility regimes above, and compare it to $\kappa_T/24$, the
curvature obtained from the Edgeworth expansion (\ref{eq:smile1}).  Here again
the results are surprising: we find that the ``fair'' curvature $\gamma_T$ of
the smile around the money is close to zero, both for high and low
volatilities. This is very different from what the empirical kurtosis of the distribution
suggests, especially in the high vol regime where the kurtosis of returns is
empirically very substantial, in agreement with the fact, noted above, that the
distribution of returns exhibits power-law tails. \footnote{The kurtosis might even 
be mathematically divergent.} This is a striking
illustration of the misleading character of the Edgeworth expansion
(\ref{eq:smile1}): the around-the-money smile of index options should be close
to a straight line with very little curvature, as indeed often seen on implied 
smiles, whereas the Edgeworth expansion would predict highly curved smiles. 

\begin{figure}[tb]
    \includegraphics[width=10cm]{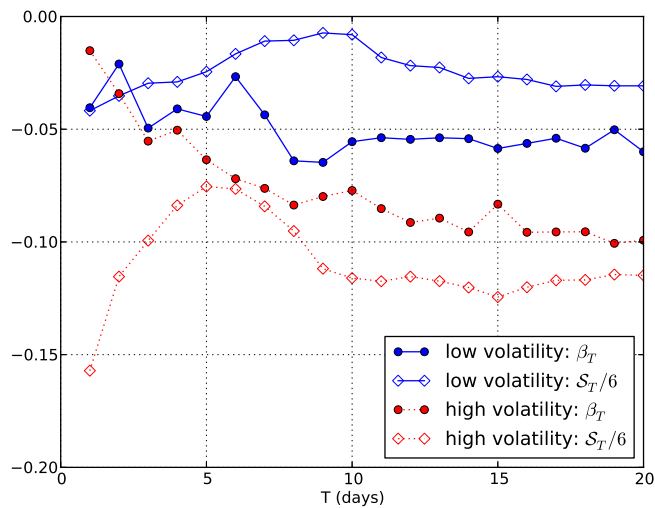}
    \caption{S\&P500 data. Skew $\beta_T$ and skewness $\mathcal{S}_T/6$ for
    different volatility regimes. Note the inversion of the order between the 
    two quantities as the volatility increases.}
\end{figure}

\begin{figure}[tb]
    \includegraphics[width=10cm]{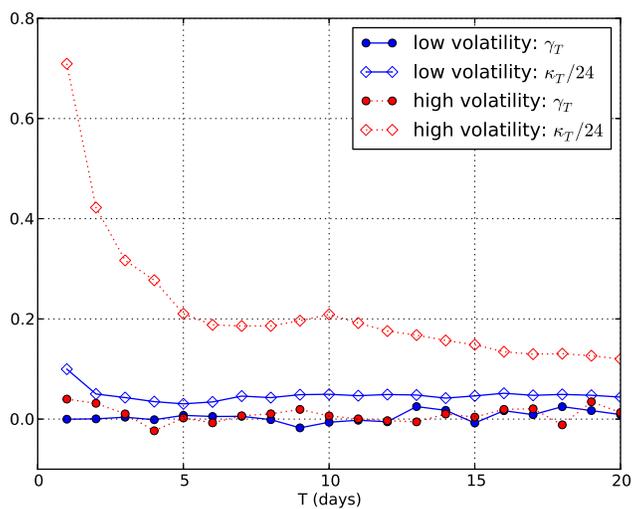}
    \caption{S\&P500 data. Curvature  $\gamma_T$ and kurtosis $\kappa_T/24$ for 
    different volatility regimes. Note that, surprisingly, the curvature of the 
    smile is found to be very small in both volatility regimes.}
\end{figure}

The results on the skew of the smile shown in Fig. 1 are clearly inconsistent 
with the relation $\beta_T = \mathcal{S}_T/6$, and cannot be
understood either within the above non-linear leverage effect with a fixed
parameter $\theta$. A way out would be to assume that $\theta$ itself depends
on the ``baseline'' volatility $\overline{\sigma}$ in Eq. (\ref{eq:non-linear}) above, 
with $\theta > 0$ during high volatility periods and $\theta < 0$ during low volatility periods. 
This could be interpreted as meaning that at high vols, a
significant down-trend is required to drive the volatility even higher, whereas at low
vols, it takes a moderate down-trend to move the volatility up. It is also
instructive to compare these results with the predictions of a popular
asymmetric GARCH model used to model the dynamics of the S\&P500 index
\cite{Berd}. Writing the returs as $\sigma_t \eta_t$ with $\eta_t \sim
{\mathcal N}(0,1)$, the dynamics of $\sigma_t$ is postulated to be:

\begin{equation}\label{eq:GARCH}
\sigma_t = \sigma (1 + \chi_t); \qquad \chi_{t+1} = 
\rho \chi_t + \nu (1+\chi_t) \left[\eta_t^2 1_{\eta_t < 0} - \frac{1}{2} \right].
\end{equation}

We have run our HMC on synthetic time series generated using the asymmetric
GARCH (GAARCH) model with parameters $\rho=0.9$ and $\nu=0.1$, corresponding to
a ``memory time'' of the volatility equal to 10 days.  One can of course now
generate as many trajectories as needed to get good statistics. The results for
$\beta_T$ and $\gamma_T$ together with their corresponding cumulant expansion
expressions are reported in Fig. 3 for the high volatility regime. We observe in
this case a trend similar to the S\&P500 data: the skewness $|\mathcal{S}_T|/6$
exceeds $|\beta_T|$ in absolute value and $\kappa_T / 24$ is systematically
larger than $\gamma_T$, although the latter is significantly different from
zero. However, for low volatilities (not shown) the behavior of the skew is
qualitativaly different from the S\&P500 data. In that case $\beta_T$ and
$\mathcal{S}_T/6$ are very close (except for very small T), but with still
$|\mathcal{S}_T|/6 > |\beta_T|$. In other words, the inversion observed on
empirical data for low vols cannot be reproduced within the GAARCH model. That
$|\mathcal{S}_T|/6 > |\beta_T|$ can in fact be checked analytically in a small
$\nu$ expansion. Note this inequality implies that the ``Skew Stickiness Ratio'' introduced by Bergomi
in \cite{BergomiIV} is expected to be smaller than 2 at small maturities.

\begin{figure}[tb]
    \includegraphics[width=10cm]{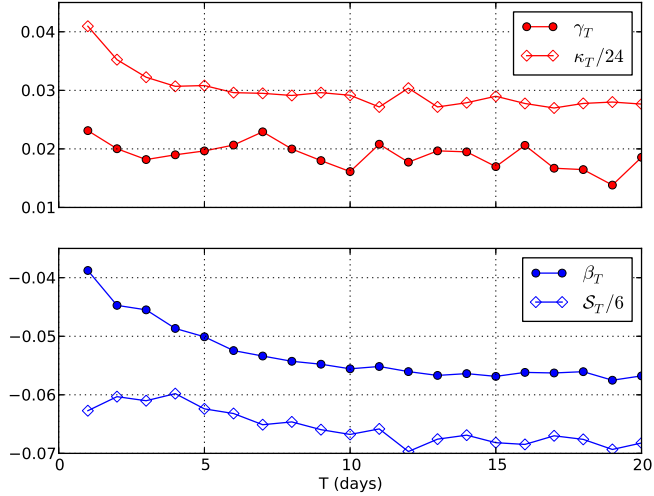}
    \caption{GAARCH synthetic data from equation (\ref{eq:GARCH}), with $\rho=0.9$, $\nu=0.1$, and 
    $10^6$ Monte Carlo paths for high volatility regime. Upper panel: curvature  $\gamma_T$  and kurtosis $\kappa_T/24$. 
    Lower panel: skew $\beta_T$ and skewness $\mathcal{S}_T/6$.}
\end{figure}

As a conclusion, we have set up a new, exact and transparent expansion for
option smiles, which lends itself both to analytical approximation and,
perhaps more importantly, to congenial numerical treatments. We have shown
that the skew and the curvature of the smile can be computed as exotic options,
for which the Hedged Monte Carlo method is particularly well suited. When
applied to options on the S\&P index, we have found that the skew and the
curvature of the smile are very poorly reproduced by the standard Edgeworth
(cumulant) expansion, or by the Bergomi-Guyon vol-of-vol expansion (at least to lowest order). 
Most notably, the relation between the skew and the skewness is inverted at small
and large vols, a feature that none of the models studied so far is able to
reproduce.  We have argued that some coupling between the leverage effect and the volatility 
level is needed to capture such non-trivial statistical features. 
Finally, the around-the-money curvature of the smile is found to
be very small, in stark contrast with the highly kurtic nature of the returns. 
This would also require some more detailed understanding. 
We hope that the ideas and methods presented in this paper are of general
interest, and provide a fruitful framework to interpret the information provided by  option
markets.

\vspace{0.5cm}

We thank Marc Potters and Arthur Berd for many stimulating discussions. We are indebted to Julien Meltz who was involved in the earlier stages of this
project. Finally, we thank L. Bergomi, J. Guyon, J. Gatheral and V. Kapoor for useful comments on the manuscript. 

\section*{Technical Appendix}

\subsection*{Proof of the smile formula}

We denote the price $S_T=S_0 (1+r_T)$. We suppose that $r_T$ is centered and that $\sigma^2 T=E[r_T^2]$. We set $u_T=r_T/ \sigma \sqrt{T}$.
We have:
\begin{align*}
E[(S_T -K)_{+}] & = S_0 E[( r_T -(\frac{K}{S_0}-1) )_{+}]  =  S_0 \int_{\frac{K-S_0}{S_0}}^{\infty} P( r_T >  x ) dx  \\
& =  S_0 \sigma \sqrt{T} \int_{\mathcal{M}}^{\infty} P( u_T>  u ) du  \\
\end{align*}
where we introduce the renormalized moneyness $\mathcal{M}=\frac{K-S_0}{\sigma \sqrt{T} S_0}$. A direct expansion  in $\mathcal{M}$ leads to:
\begin{equation*}  
E[  ( S_T -K )_{+} ]  \approx  S_0 \sigma \sqrt{T} (   \frac{E[  | u_T | ]}{2}- \mathcal{M} P( u_T>  0)+\frac{\mathcal{M}^2}{2}p_T(0)  )
\end{equation*}
In the Gaussian case, we get therefore:
\begin{equation*}
E[  ( S_T -K )_{+} ]  \approx S_0 \sigma_{BS} \sqrt{T} (   \frac{1}{\sqrt{2\pi }}- \frac{\mathcal{M}_G}{2}+\frac{\mathcal{M}_G^2}{2}\frac{1}{\sqrt{2\pi }}  )
\end{equation*}
where $\mathcal{M}_G=\frac{K-S_0}{\sigma_{BS} \sqrt{T} S_0}$ is the Gaussian moneyness.

We make now make the assumption:
\begin{equation}\label{eq:smileexp}
\sigma_{BS}= \sigma ( \alpha' + \beta'  \mathcal{M}_G + \gamma' \mathcal{M}_G^2  ),
\end{equation}
so by definition $\mathcal{M}=\mathcal{M}_G (\alpha' + \beta' \mathcal{M}_G + \gamma' \mathcal{M}_G^2  )$. The smile formula corresponds to the following identity:
\begin{equation*}
\sigma_{BS} (   \frac{1}{\sqrt{2\pi }}- \frac{\mathcal{M}_G}{2}+\frac{\mathcal{M}_G^2}{2}\frac{1}{\sqrt{2\pi }}  )= \sigma (   \frac{E[  | u_T | ]}{2}- \mathcal{M} P( u_T>  0)+\frac{\mathcal{M}^2}{2}p_T(0)  )
\end{equation*}
Therefore, we get the following equation (to the order $\mathcal{M}_G^2$):
\begin{equation*}
( \alpha' + \beta' \mathcal{M}_G + \gamma' \mathcal{M}_G^2  )(   \frac{1}{\sqrt{2\pi }}- \frac{\mathcal{M}_G}{2}+\frac{\mathcal{M}_G^2}{2}\frac{1}{\sqrt{2\pi }}  ) = 
(   \frac{E[  | u_T | ]}{2}- (\alpha' \mathcal{M}_G+\beta' \mathcal{M}_G^2 ) P( u_T>  0)+\frac{\alpha'^2\mathcal{M}_G^2}{2}p_T(0)  ).  
\end{equation*}
which finally lead to the following identification:
\begin{equation*}
\left\{\begin{array}{l}
\alpha'=\sqrt{\frac{\pi}{2}}E[  | u_T | ]\\
\beta'= \pi E[  | u_T | ] (\frac{1}{2}-P( u_T>  0))\\
\gamma'= \sqrt{2\pi}(   \frac{\alpha'^2p_T(0)}{2}  -\frac{\alpha'}{2\sqrt{2\pi }}  +\pi E[  | u_T | ]  (\frac{1}{2}-P( u_T>  0)) ^2 ) \\
\end{array}\right.
\end{equation*}
Switching back to standard moneyness and using $\sigma_{BS}= \sigma ( \alpha + \beta \mathcal{M} + \gamma \mathcal{M}^2  )$ finally leads to:
$$\left\{\begin{array}{l}
\alpha=\sqrt{\frac{\pi}{2}}E[  | u_t | ]\\
\beta= \frac{\beta'}{\alpha}= \sqrt{2\pi}(\frac{1}{2}-P( u_t>  0))\\
\gamma=\frac{\gamma'}{\alpha^2}-\frac{\beta'^2}{\alpha^3}=  \sqrt{2\pi}(   \frac{p_T(0)}{2}  -\frac{1}{2\alpha \sqrt{2\pi }})\\
\end{array}\right.$$

\subsection*{The Edgeworth expansion}

Assuming $\mathcal{S}_T, \kappa_T\ll1$, the Edgeworth expansion reads: 
\begin{equation*}
P(\frac{r_T}{\sigma \sqrt{T}}>x)-\overline{N}(x) \approx \frac{\mathcal{S}}{6}  N^{(3)}(x) -\frac{\kappa}{24} N^{(4)}(x).
\end{equation*}
where $N(x)=\int_{-\infty}^{x}\frac{e^{-t^2/2}}{\sqrt{2 \pi}}dt$,  $\overline{N}=1-N$ and $N^{(n)}$ refers to the $n$th derivative of $N$.
This leads to the following approximations: 
\begin{equation*}
\left\{\begin{array}{l}
\alpha \approx 1-\frac{\kappa_T}{24}\\
\frac{1}{2}-P( u_T>  0) \approx -M_T p_T(0) \approx  -\frac{M_T}{\sqrt{2 \pi}} \ll1\\
p_T(0)-\frac{1}{\sqrt{2\pi }} \approx \frac{1}{\sqrt{2\pi }} \frac{\kappa_T}{8}\ll1 \\
\end{array}\right.
\end{equation*}
where $M_T$ is the median of $u_T$. 
With these assumptions, we get the following expressions for $\alpha,\beta,\gamma$:
\begin{equation*}
\left\{\begin{array}{l}
\alpha \approx 1-\frac{\kappa_T}{24}\\
\beta \approx -M_T\\
\gamma \approx  \frac{\sqrt{2\pi}}{2}(\frac{1}{\sqrt{2\pi }}(1+\frac{\kappa_T}{8})-\frac{1}{\sqrt{2\pi }}(1+\frac{\kappa_T}{24}))\approx \frac{\kappa_T}{24}\\
\end{array}\right.
\end{equation*}
Within the aforementioned approximation, both smile formulas coincide since we know that the Edgeworth expansion also leads to:
\begin{equation*} 
\mathcal{S}_T \approx - 6 M_T , \quad \kappa_T \approx 24(1-\sqrt{\frac{\pi}{2}} E[|u_T|])
\end{equation*}

\bibliography{references}{}

\end{document}